\documentstyle[12pt,fleqn]{article}
\def\thesection{}
\def\bfx{{\bf x}}
\def\bfv{{\bf v}}

\def\smth#1{{\langle{#1}\rangle}}
\def\suma#1{\lbrack\!\lbrack{#1}\rbrack\!\rbrack}
\def\dt{\Delta t}
\def\dx{\Delta x}
\begin{document}
\begin{titlepage}
\vspace*{-62pt}
\begin{flushright}
{\footnotesize
PSU-ASTR0 94/2-1 \\
(February 1994)}
\end{flushright}
\renewcommand{\thefootnote}{\fnsymbol{footnote}}
\begin{center}
{\Large \bf Smoothed Particle Interpolation}\\

\vspace{0.6cm}
\normalsize

Pablo Laguna\footnote{e-mail: pablo@astro.psu.edu}\\
{\em Department of Astronomy and Astrophysics\\
     and\\
     Center for Gravitational Physics and Geometry\\
     The Pennsylvania State University\\
     University Park, PA 16802}\\

\end{center}

\vspace*{12pt}

\begin{quote}
\hspace*{2em}Smoothed particle hydrodynamics (SPH) discretization
techniques are generalized to develop a method, smoothed particle
interpolation (SPI), for solving initial value problems
of systems of non-hydrodynamical nature.
Under this approach,  SPH is viewed as strictly an
interpolation scheme and, as such, suitable for solving
general hyperbolic and parabolic equations.
The SPI method is tested on (1) the wave equation
with inhomogeneous sound speed and (2) Burgers equation.
The efficiency of SPI is studied by comparing SPI solutions to
those obtained with standard finite difference methods.
It is shown that the power of SPI arises when
the smoothing particles are free to move.

\bigskip
\noindent{\it Subject headings:} hydrodynamics --
methods: numerical

\vspace*{12pt}

\small
\end{quote}
\renewcommand{\thefootnote}{\arabic{footnote}}
\addtocounter{footnote}{-4}

\end{titlepage}

\thesection{\centerline{\bf 1. INTRODUCTION}}
\setcounter{section}{1}
\setcounter{equation}{0}
\vspace{12pt}

Perhaps one of the most powerful aspects in computational studies
is the development of a fully adaptive scheme
for solving partial differential equations.
Without adaptive numerical schemes,
current large memory and fast supercomputers
are not capable yet of obtaining highly accurate solutions to physically
interesting three-dimensional problems. For this reason,
an optimal management of computational resources has been
one of the driving
forces in the effort to achieve both temporal and
spatial adaptability as a way of obtaining a higher degree of
accuracy.

The goal for reaching optimal temporal adaptability consists of
designing algorithms that evolve each region of the computational domain
with time-steps according to their local time-scales.
In particle methods, local time-step algorithms are aimed at
subsets of particles. The main difficulty here arises
in designing a time evolution algorithm compatible with
the neighbor-finding scheme (tree-code).
In the case of finite differencing, the target of adaptability
are subgrids. Each subgrid, for instance, is evolved according to its own
Courant condition. Special attention must be paid at the subgrid
boundaries where gridpoints in coarser meshes are not updated
as often as in the finer meshes.
Other approaches to temporal adaptability exploit
multiple time-stepping applied at the equation
instead of the domain level.
These methods rely on the possibility of identifying
in the evolution equations
the slowly varying terms that require time-step $\dt$
in contrast to rapidly varying terms with time-step $\delta t < \dt$.
The slowly varying terms are then computed once every
large time-step $\dt$, and the rapidly
varying terms are subcycled $\dt / \delta t$
times for each large time-step $\dt$.
An example of this approach is called
acoustic-subcycling, which has successfully improved the efficiency
of compressible flow codes (O'Rourke \& Amsden 1986) when applied to low
Mach number problems.

Technically speaking, spatial adaptability exhibits a higher degree of
difficulty than temporal one. For finite difference methods,
this translates into designing an efficient remeshing algorithm.
Spatially adaptive finite difference methods for hyperbolic equations have
been proposed by Berger and Oliger (1984) using multiple component grids.
This approach was the keystone for Choptuik's (1993) finding of
scaling and critical behavior in the gravitational collapse
of a scalar field.
Adaptive finite element methods, which have been mainly used
for elliptic problems, have been also applied to
hyperbolic equations (Davis \& Flaherty 1982).
Finally, particle methods have been explicitly designed
to exploit adaptability; a set of moving points carry the information
of the dynamics of the  system. Smoothed particle hydrodynamics (SPH)
(for a review, see Benz 1990) is an
example of such methods and constitutes the focus of this paper.

SPH has become a remarkably robust hydrodynamic method, in particular
for astrophysical systems where vast range of scales are present. It
has been successfully applied to
stellar encounters(Davies, Benz, \& Hills 1991), i
galaxy formation (Hernquist \& Katz 1989),
tidal disruptions (Evans \& Kochanek 1989; Laguna, Miller, Zurek, \& Davies
1993),
and more recently to problems outside the
astrophysical domain, such as hypervelocity impacts.
SPH has also been reformulated to describe relativistic fluids in
curved space-times (Laguna, Miller \& Zurek 1993).
SPH has been also used in combination with other methods in studying
mixed systems where additional physics is required
(hydrodynamics + magnetism, hydrodynamics + gravity,
hydrodynamics + collisionless matter, etc.). In most of these cases, SPH
is only applied to the hydrodynamic equations, and other numerical methods
are used in the remaining set of equations.
Mann (1993) has solved the relativistic, self-gravitating spherical collapse
using finite element methods to compute the gravitational potentials
and SPH for the hydrodynamics equations.
In the study of large scale structure formation, SPH is combined with
$N$-body methods to simulate hybrid systems of Hot (SPH) + Cold ($N$-body)
Dark Matter. There are, however, examples of studies in which,
still in the presence of fluids, SPH has been applied to non-hydrodynamical
equations. For instance, in the low frequency and infinite conductivity limit
Stellingwerf and Peterkin (1989) have developed a simple and elegant
method for smoothed particled magnetohydrodynamics.

The main goal of this paper is to address the following question:
Can the success of SPH be extended to purely non-hydrodynamical problems?
In other words, can one take advantage of SPH gridless nature of
computing spatial derivatives and move particles with arbitrary ``grid"
velocity?
We present one-dimensional results that point in the direction
of an affirmative answer to this question.
In Section 2 of the paper, we present the method for applying SPH
discretization techniques to general hyperbolic and parabolic equations,
giving special attention to truncation errors.
Section 3 contains one-dimensional
tests consisting of solutions to
the wave equation with inhomogeneous sound speed;
for these tests the particles are ``fixed" in time (Eulerian framework).
In Section 4, we repeat the tests of Section 3, but now
we allow for particle motion (Lagrangian framework).
Solutions to Burgers equation using SPH-type methods are obtained in Section 5.
Finally, discussion and conclusions are given in Section 5.

\vspace{24pt}
\thesection{\centerline{\bf 2. SPI: SMOOTHED PARTICLE INTERPOLATION}}
\setcounter{section}{2}
\setcounter{equation}{0}
\vspace{12pt}

The key ingredients of SPH are:
(1) the fluid elements of the system are represented by particles,
(2) the dynamics of the particles is governed by the
hydrodynamic equations written in the Lagrangian form, and (3) the
spatial derivatives at the particle positions are calculated via
an interpolation kernel. Given a physical function
$f(\bfx)$ (pressure, density, etc.), its mean smoothed
value can be obtained from
\begin{equation}
\label{smooth}
\langle f(\bfx)\rangle  \equiv \int W(\bfx,\bfx';h)f(\bfx') d^3x',
\end{equation}
where $W(\bfx,\bfx';h)$ is the kernel and $h$ the smoothing length.
The kernel $W(\bfx,\bfx';h)$ has compact support in a region of $O(h)$,
and for hydrodynamic calculations the kernel is assumed to be of class
$C^1$, i.e., its first derivatives exist and are continuous.
One can show that this smoothing procedure is second-order accurate; that is,
$\langle f(\bfx)\rangle = f(\bfx) + O(h^2).$
In particular, for the case of an spherically symmetric kernel the
truncation errors due to the
smoothing are
\begin{equation}
\label{smooth_error}
\langle f(\bfx)\rangle = f(\bfx) + \alpha\, h^2\, \Delta f + O(h^3),
\end{equation}
where
\begin{equation}
\alpha = \int W(|\bfx-\bfx'|;h) |\bfx-\bfx'|^2 h^3 d^3x'
\end{equation}
is independent of $h$; in the one-dimensional case, $\alpha=1/4$.
The consistency requirement that
$\langle f(\bfx)\rangle \rightarrow f(\bfx)$ as
$h \rightarrow 0$, imposes the following normalization condition on $W$:
\begin{equation}
\label{norm}
\int W(\bfx,\bfx';h) d^3x = 1.
\end{equation}
The smoothing procedure (\ref{smooth}) provides a natural
prescription for calculating spatial derivatives at the
particle locations. One has that, when ignoring surface terms,
integration by parts of
\begin{equation}
\langle \nabla f(\bfx)\rangle  =
\int W(\bfx,\bfx';h)\nabla' f(\bfx') d^3x'
\end{equation}
yields,
\begin{equation}
\label{grad1}
\langle \nabla f(\bfx)\rangle  =
\int f(\bfx')\nabla W(\bfx,\bfx';h) d^3x',
\end{equation}
where $\nabla'$ and $\nabla$ are gradients with respect to
$\bfx'$ and $\bfx$, respectively.
Equation~(\ref{grad1}) is just the statement that
$
\langle \nabla f(\bfx)\rangle  = \nabla\langle f(\bfx)\rangle.
$
Similarly, it is not difficult to show that
for a vector field $\bfv(\bfx)$,
the smoothed value of its divergence is given by
$
\langle \nabla\cdot \bfv(\bfx)\rangle
= \nabla\cdot\langle \bfv(\bfx)\rangle,
$
or equivalently by
\begin{equation}
\label{grad2}
\langle \nabla\cdot\bfv(\bfx)\rangle  =
\int\bfv(\bfx')\cdot\nabla W(\bfx,\bfx';h) d^3x'.
\end{equation}
Finally, the smoothing of $\nabla^2f$
is obtained by successively applying
properties $\smth{\nabla\cdot\bfv(\bfx)} = \nabla\cdot\smth{\bfv(\bfx)}$
and $\smth{\nabla f(\bfx)}=\nabla\smth{f(\bfx)}$. One gets that
$\smth{\nabla^2f} = \nabla^2\smth{f}$, which
has as integral representation
\begin{equation}
\label{grad3}
\smth{\nabla^2 f(\bfx_a)}  =
\int f(\bfx') \nabla^2 W(\bfx,\bfx';h) d^3x'.
\end{equation}

One can show that the second-order accurate nature of
the smoothing procedure is preserved under differentiation.
The differentiation rules (\ref{grad1}) and (\ref{grad2})
to compute $\langle \nabla f(\bfx)\rangle$
and $\langle \nabla\cdot \bfv(\bfx)\rangle$, together with
definition (\ref{smooth}), is all that is needed to
derive the SPH equations in their integral form (Benz 1990).

The next step is to obtain the discrete version of
Eqs.~(\ref{smooth}) and (\ref{grad1})-(\ref{grad3});
that is, a discrete representation of the integrals in those
equations. In order to accomplish this, one defines first a
number density distribution
\begin{equation}
\label{distri}
n(\bfx) \equiv \sum_{a=1}^N \delta(\bfx-\bfx_a),
\end{equation}
with $\lbrace \bfx_a \rbrace_{a=1,..,N}$
the collection of $N$-points (particles)
where the physical functions are known.
The integrals in Eqs.~(\ref{smooth}) and
(\ref{grad1})-(\ref{grad3}) are evaluated by
multiplying the integrands by $n(\bfx)/\langle n(\bfx)\rangle = 1 + O(h^2)$,
yielding
\begin{equation}
\label{sphform1}
\suma{f(\bfx_a)}  =
\sum_{b=1}^N \frac{f(\bfx_b)}{\langle n(\bfx_b)\rangle }
W(\bfx_a,\bfx_b;h) \,,
\end{equation}
\begin{equation}
\label{sphform2}
\suma{\nabla f(\bfx_a)}  =
\sum_{b=1}^N \frac{f(\bfx_b)}{\langle n(\bfx_b)\rangle }
\nabla_{\bfx_a} W(\bfx_a,\bfx_b;h)  \,,
\end{equation}
\begin{equation}
\label{sphform3}
\suma{\nabla\cdot \bfv(\bfx_a)}  =
\sum_{b=1}^N \frac{\bfv(\bfx_b)}{\langle n(\bfx_b)\rangle }
\cdot\nabla_{\bfx_a} W(\bfx_a,\bfx_b;h) \,,
\end{equation}
and
\begin{equation}
\label{sphform4}
\suma{\nabla^2 f(\bfx_a)}  =
\sum_{b=1}^N \frac{f(\bfx_b)}{\langle n(\bfx_b)\rangle }
\nabla^2_{\bfx_a} W(\bfx_a,\bfx_b;h).
\end{equation}
respectively.
The SPH equations, in their sum representation,
can be directly derived by substitution of
(\ref{sphform1})-(\ref{sphform3}) into
the hydrodynamic conservation laws written in a Lagrangian form (Benz 1990).

Higher order derivatives can be obtained following a procedure
similar to that used to derive Eqs.~(\ref{sphform1})-(\ref{sphform2}).
In principle, the only requirement needed is that a
kernel of class $C^n$ is used in the smoothing of
a $n$-order derivative. For instance,
the smoothed approximation to the Laplacian operator,
given by Eq.~(\ref{sphform4}) requires a $C^2$ kernel, so the popular
spline kernel (Monaghan \& Lattanzio 1985) used in SPH fails because of its
discontinuous second derivatives.
For this reason, we will use a gaussian kernel
\begin{equation}
\label{gauss}
W(r,h) = \frac{1}{\pi^{d/2}h^d}\cases{
\exp{(-u^2)}&($u<u_o$)\cr
\noalign{\vskip2pt}
0&($u>u_o$)\cr}
\end{equation}
where $u \equiv |\bfx_a-\bfx_b|/h$, $d = 1,2,3 $
denotes the dimension of the problem, and
$u_o$ the kernel's compact support radius (Figure~1).
As we will see below, a disadvantage of this kernel
is that it needs to be defined over a larger compact support,
typically $u_o \sim 5$.
In contrast, the spline's compact support radius is
by definition $u_o = 2$;
hence, in general, using a gaussian kernel requires a
larger number of neighbors
to achieve similar accuracy as with the spline kernel.

It is important to notice that we have distinguished the
smoothing procedure $\langle\,\,\rangle$ from the
discrete smoothing procedure $\suma{\,\,} $ to emphasize that
there are in general two sources of truncation errors in approximating
a function for its smoothed-discrete value; one error,
the smoothing-error ($\smth{\epsilon}$),
arises from the smoothing procedure itself (see Eq.~(\ref{smooth_error})),
and a second error, the sum-error ($\suma{\epsilon}$),
is generated from the discretization of
the integrals; that is
\begin{equation}
\label{all_error}
\sum_{b=1}^N \frac{f(x_b)}{n(x_b)}\frac{d^n}{dx^n}W(x,x_b;h) =
\frac{d^n}{dx^n}f(x)+\smth{\epsilon}+\suma{\epsilon}.
\end{equation}
It is straightforward to show that for the one-dimensional gaussian kernel,
the generalization of the
smoothing-error (\ref{smooth_error}) to $n$-order derivatives is
\begin{equation}
\label{smoothing-error}
\smth{\epsilon} = \frac{h^2}{4}\,\frac{d^{n+2}}{dx^{n+2}}f(x)+O(h^3).
\end{equation}

The distribution of the particles plays a fundamental role in
computing the truncation sum-error. If the particles have
a chaotic distribution, the approximations
(\ref{sphform1})-(\ref{sphform4})
would be Monte Carlo estimates of the integrals in
Eqs.~(\ref{smooth}) and (\ref{grad1})-(\ref{grad3}), respectively.
However, in SPH the particles do not follow a chaotic position
distribution; they are not completely independent.
On the other hand, they are not, in general, distributed
in a regular configuration either.
Nonetheless, for estimating sum-errors, we are going to
assume a regular distribution of particles because it is in this case
for which the minimum truncation sum-error is obtained.
This will guarantee that features arising from the sum-error in regular
particle distributions will also show up in more realistic configurations.
For simplicity, we will only consider here the one-dimensional case.

Following a procedure similar to that of deriving truncation errors
in standard quadrature formulas, it is not difficult to show that
the leading truncation sum-error in (\ref{all_error}) is given by
\begin{equation}
\label{sum_error}
\suma{\epsilon}=
h^{-n}\biggl(\frac{\dx}{h}\biggr)^2\,f(x)\frac{d^{n+2}}{du^{n+2}}w(u)
+O((\dx/h)^4\,h^{-n}),
\end{equation}
where $\dx$ is the particle interseparation
and $w(u) \equiv h W(u,h)$ only depends on $u$. Furthermore,
$h/\dx = N_{1/2}/u_o$, where $u_o$ is the compact support radius
and $N_{1/2}$ denotes the number of neighbors
of a particle at each of its ``sides" (Figure~1).
It is important to notice from (\ref{sum_error}) that
in the case of ``pure" smoothing ($n=0$), the sum-error
only depends on $h/\dx \propto N_{1/2}$, the number of neighbors.

Convergence tests in SPH have been usually done by doubling the
total number of particles; for our one-dimensional case,
this implies $\dx^{(k)} = 2^{-k}\dx_o$ with $k = 0,1,...$ and
$\dx_o$ the spacing of the coarsest mesh.
Little mention has been given, though, to
how the number of neighbors is handled, i.e.,
the $h/\dx \propto N_{1/2} \rightarrow \infty $ limit.
One has to remember that there are two competing errors in going from
the continuum expressions to the discrete, smoothed approximations.
The smoothing-error has a discretization scale given by the
smoothing length $h$, see Eq.~(\ref{smoothing-error}),
and the sum-error depends, on the other hand, not only
on the number of neighbors $h/\dx \propto N_{1/2}$
within the compact support of the kernel, but also on the smoothing
length $h$ for terms involving differentiation,
see Eq.~(\ref{sum_error}) with $n>0$.
Convergence tests have to satisfy the condition
$h^{-n}(\dx/h)^2 \rightarrow 0$ as
$\dx$ and $h \rightarrow 0$. To illustrate this, we have
computed the smoothed values of an analytic function
and its smoothed first and second derivatives for different
resolutions; that is, $\dx^{(k)} = 2^{-k}\dx_o$ and
$h^{(l)} = 2^{-l}h_o$ with $k,l = 1,..,4$.
Figure~2 shows the results using $f(x) = \sin(x)$.
The lines join points of constant $\dx$.
As expected, the top two plots show the second
order convergence rate of $\suma{f(x)}$.
A similar second-order behavior is obtained
for $\suma{f'(x)}$ and $\suma{f''(x)}$
if $N_{1/2} > 8$.
However, for $N_{1/2} < 8$,
there seems to be an ill defined limit for
$\suma{f'(x)}$ and $\suma{f''(x)}$
when $h \rightarrow 0$; there the errors grow as $h \rightarrow 0$.
As mentioned before, the answer to this puzzle is in how the limit
$h^{-n}(\dx/h)^2$ is taken.
We see that in particular the limit used in Fig. 2,
$\dx, h \rightarrow 0$ with $h/\dx \propto N_{1/2}$ constant,
will break down for $N_{1/2} < 8$. Of course in practice, it is impossible
to take the limit to the continuum; however, it is
important to have, at a given resolution scale, enough number
of neighbors, so the smoothing procedure falls within the second-order regime
of the
truncation errors. In our one-dimensional case, this means
to have $N_{1/2} > 8$.

Finally, in SPH the particles are identified with fluid elements;
hence, their dynamics is governed by the
hydrodynamic equations. In generalizing SPH methods
beyond the scope of hydrodynamics, we view SPH as strictly an
interpolation scheme (Smoothed Particle Interpolation: SPI)
and, as such, used only to
compute spatial gradients without the need of introducing
a grid. The particles could be either ``fixed" in space,
although not necessarily uniformly distributed, or free to move.
The latter case will represent a truly adaptive implementation.
For the moving particles (Lagrangian) case, a
rule that dictates motion of the particles should be provided.
In principle, the rule for moving particles can be arbitrarily chosen;
however, it is not clear at this stage whether non-single valued velocity
fields are recommended since they could possibly lead to difficulties
such as the one arising in hydrodynamic calculations
when particle penetration occurs (Monaghan 1989).

\vspace{24pt}
\thesection{\centerline{\bf 3. EULERIAN SPI}}
\setcounter{section}{3}
\setcounter{equation}{0}
\vspace{12pt}

In this section we will concentrate our attention on testing only
the interpolating features of SPI.
Hence, the interpolating particles will remain fixed in their original
positions through out the evolution;
the adaptive (moving particles) properties of SPI will be addressed
in the next section.
For these fixed particle tests,
we start by considering the wave equation
\begin{equation}
\label{hyper}
\partial^2_t\phi(\bfx)-c^2(\bfx)\nabla^2\phi(\bfx)=0,
\end{equation}
where $\partial^2_t \equiv \frac{\partial^2}{\partial t^2}$.
We define the ``momentum" variable $\chi \equiv \partial_t\phi$ , and
rewrite Eq.~(\ref{hyper}) as
\begin{equation}
\label{sound1}
\partial_t\phi=\chi.
\end{equation}
and
\begin{equation}
\label{sound2}
\partial_t\chi=c^2\nabla^2\phi
\end{equation}

We now apply smoothing to
(\ref{sound1}) and (\ref{sound2}). It is important
to keep in mind that
the smoothing properties act only on spatial directions; hence,
temporal differentiation and smoothing commute,
$\smth{\partial_t f} = \partial_t\smth{f}$.
For temporal discretization, we use standard finite difference schemes.
{}From substituting (\ref{sphform1}) and (\ref{sphform4})
into Eqs.~(\ref{sound1}) and (\ref{sound2}),
one obtains the discrete SPI approximations to Eqs.~(\ref{sound1})
and (\ref{sound2}) to be
\begin{equation}
\label{sound_sum1}
\frac{\phi_a^{n+1} -\phi_a^n}{\dt} = \chi_a^{n+1/2},
\end{equation}
and
\begin{equation}
\label{sound_sum2}
\frac{\chi_a^{n+3/2} - \chi_a^{n+1/2}}{\dt} =
c^2_a\sum_{b=1}^N \frac{\phi_b^{n+1}}{n_b}\nabla^2_a W_{ab},
\end{equation}
where $W_{ab} \equiv W(\bfx_a,\bfx_b;h)$.
The time step is given by $\dt = D h/max(c_s(x))$
with $D < \sqrt{3}/2$.
It is evident from these expressions that temporal integration
is performed using a leap frog method (Roache 1985).
There are two interlaced time-steps (integer- and half-steps).
The terms $\partial_t\chi$, $\phi$ and spatial derivatives of $\phi$
are placed at integer-steps; similarly,
$\partial_t\phi$, $\chi$ and spatial derivatives $\chi$
are placed at half time-steps. This temporal discretization
makes the numerical integration second-order accurate.

To investigate the effects of the numerical dispersion in SPI, we
evolved a gaussian pulse through a homogeneous medium; $c(x) = 1$ in
Eq.~(\ref{sound1}).
The initial conditions for $\phi$ and its ``momentum" $\chi$ were
\begin{equation}
\label{pulse_form1}
\phi(x) = \phi_o\exp{[-k^2(x - t)^2]}
\end{equation}
and
\begin{equation}
\label{pulse_form2}
\chi(x) = 2\,k^2(x - t)\phi(x),
\end{equation}
respectively. Solution errors are presented as cumulative relative errors;
that is,
\begin{equation}
\label{cumm_error}
\epsilon = \biggl(\frac{\sum_{a=1}^N (\widetilde\phi_a-\phi_a)^2}
                 {\sum_{a=1}^N \widetilde\phi_a^2}\biggr)^{1/2},
\end{equation}
where $\phi$ and $\widetilde\phi$ are the numerical and analytical solutions,
respectively.
Figure~(3.a) shows the solution error as a function of the
resolution scale following a displacement of four gaussian width-lengths.
The second order convergence rate implied by
Eq.~(\ref{smooth_error}) is evident from this figure.
For comparison, in Fig.~(3.b) we plot the difference between
SPI and FD solution errors;
the error differences are plotted as a function of resolution scale.
The resolution scale is defined
as the discretization scale (particle interseparation in SPI or
mesh spacing in FD) for which both methods, SPI and FD,
yield the same absolute truncation error. It is not difficult to
show by substituting (\ref{gauss}) into Eq.~(\ref{smooth_error}),
that for uniformly spaced particles given a grid spacing $\dx(FD)$ in FD,
a smoothing length of $h = \dx(FD) / \sqrt{3}$ yields the same
SPI and FD truncation errors. Furthermore, one also gets that the ratio of
SPI particle spacing to FD grid spacing is given by
$\dx(FD)/\dx(SPI) = \sqrt{3} N_{1/2}/u_o$, where $N_{1/2}$ is the number
of neighbors at either side of each particle and $u_o$ the radius of the
support of the kernel in smoothing length units. As we mentioned before,
typically
$u_o \sim 5$ and $N_{1/2} \sim 8$; therefore, $\dx(FD)/\dx(SPI) \sim 2.8$.
That is, for each FD grid point, we need almost three SPI particles to achieve
the same absolute solution error. However, it is important to emphasize that
both methods have the same convergence rate, namely $O(h^2)$ for SPI and
$O(\dx^2)$ for FD.

The next test constists of a one-dimensional gaussian pulse
propagating through a medium
in which the sound speed has a jump at the origin. That is,
\begin{equation}
\label{sound_jump}
c(x) = c_o \frac{[3+\tanh{(x/\sigma)}]}{2}
\end{equation}
where, for simplicity, we use $c_o = 1$, and $\sigma$ is
chosen such that the jump is smooth
over a few particles. The pulse has a gaussian
profile as before, see Eq.~(\ref{pulse_form1}),
and is initially in the region $x < 0$ where the sound speed is unity.
Figure~(4) shows four snapshots of the evolution.
As expected, the pulse scatters at the point
where the speed of sound has a jump.
Two pulses emerge, one propagating in the same direction as
the original pulse and a second one in the opposite direction.
The calculation was repeated using a
FD method; the bottom plot shows the difference between the SPI and FD solution
for the last snapshot.

\vspace{24pt}
\thesection{\centerline{\bf 4. LAGRANGIAN SPI}}
\setcounter{section}{4}
\setcounter{equation}{0}
\vspace{12pt}

We allow now for particle motions.
However, for simplicity, we restrict ourselves to
motion of particles in a constant velocity field.
We start by rewriting equations (\ref{sound1}) and (\ref{sound2})
in the frame of reference $(t,\bfx)$ comoving with the particles.
The coordinate transformations between the comoving and laboratory frame
$(t',\bfx')$ are $t = t'$ and $\bfx = \bfx' - \bfv t'$, with
$\bfv$ the particle velocity.
To avoid complicated notation,
we have redefined the prime coordinate system
as the laboratory frame.
In these new coordinates, Eqs.~(\ref{sound1}) and (\ref{sound2}) take the form
\begin{equation}
\label{sound_v1}
\partial_t\chi-\bfv\cdot\nabla\chi=c^2\nabla^2\phi
\end{equation}
and
\begin{equation}
\label{sound_v2}
\partial_t\phi-\bfv\cdot\nabla\phi=\chi,
\end{equation}
respectively.
One immediately sees that
a description from a frame of reference comoving with the particles
introduces transport-like terms in the equations; that is, terms
of the form $\bfv\cdot\nabla$ but with the ``wrong" transport velocity sign.
These transport-like terms require the same careful consideration
to prevent the development of instabilities
as that given to transport terms in the density, energy and momentum
equations in fluid dynamics.
In addition to developing a stable form of the equations,
a ``correct" centering of each term in the equations must be
implemented in order to preserve the second-order nature of the scheme.

Let us look first at Eq.~(\ref{sound_v2}).
As mentioned in the previous section, in a leap-frog integration method,
spatial derivatives of $\nabla\phi$ are placed at integer-steps; on the
other hand, $\partial_t \phi$ and $\chi$ are located at half-steps. Hence,
an approximation to $\nabla\phi$ at half time-steps is needed.
We achieve this with the following second-order interpolation
\begin{equation}
\label{interpol}
\phi^{n+1/2} = \frac{3\,\phi^n-\phi^{n-1}}{2},
\end{equation}
with similar interpolation for $\chi$;
that is, it is necessary to save previous time-step values of $\phi$ and
$\chi$.

Using (\ref{sphform1}), (\ref{sphform2})
and (\ref{sphform4}), one obtains
the smoothed approximations of Eqs.~(\ref{sound_v1}) and (\ref{sound_v2}) to be
\begin{equation}
\label{sound_v_sum1}
\frac{\phi_a^{n+1}-\phi_a^n}{\dt} =
\sum_{b=1}^N \frac{\phi_b^{n+1/2}}{n_b}\bfv_a\cdot\nabla_a W_{ab}
+\beta\,v_a^2\,\dt\,\sum_{b=1}^N \frac{\phi_b^{n+1/2}}{n_b}\nabla^2_a W_{ab}
+\chi_a,
\end{equation}
and
\begin{equation}
\label{sound_v_sum2}
\frac{\chi_a^{n+3/2}-\chi_a^{n+1/2}}{\dt} =
\sum_{b=1}^N \frac{\chi_b^{n+1}}{n_b}\bfv_a\cdot\nabla_a W_{ab}
+c^2_a\sum_{b=1}^N \frac{\phi_b^{n+1}}{n_b}\nabla^2_a W_{ab},
\end{equation}
respectively.
We have added to (\ref{sound_v1}) a diffusion term,
second term in the r.h.s. of Eq.~(\ref{sound_v_sum1}), of the form
$\beta\,v^2\,\dt\,\nabla^2\phi$ with $\beta$ a dimensionless
parameter of order unity. Without this numerical dissipation the equations
are correctly centered but unstable.

We repeated the simulation in Sec. 3 of propagating
a pulse in a medium with a homogeneous speed of sound, $c(x)=1$ everywhere,
but now the particle velocity is $v=c$; the pulse remains
stationary in the particle frame.
Because of the artificial dissipation term in Eq.~(\ref{sound_v_sum1}),
the numerical solutions will drift from the analytic solutions
of the continuum wave equation.
Since we are interested for the moment in studying
the convergence rate of SPI,
we have added in Eq.~(\ref{sound_v_sum1})
a source term $-\beta\,v^2\,\dt\,\nabla^2\widetilde\phi$
with $\widetilde\phi$ the analytic solution.
With this term, Eqs.~(\ref{sound_v_sum1}) and (\ref{sound_v_sum2})
have as analytic solution the gaussian pulse (\ref{pulse_form1}).
This allows a direct estimate of the solution error (Choptuik 1986)
from Eq.~(\ref{cumm_error}).
We performed a series of runs with smoothing lengths
$h^{(k)} \equiv 2^{-k}h_o,\, k = 1,..,4$,
where $h_o$ is the smoothing length of the coarsest level.
The convergence factor of the solution error
at the $k$ refinement level is defined as (Choptuik 1991)
\begin{equation}
\label{c_factor}
\xi^{(k)} \equiv \frac{\epsilon^{(k)}}{\epsilon^{(k+1)}},
\end{equation}
where $\epsilon^{(k)}$ is the solution error
(See Eq.~\ref{cumm_error}) at level $k$.
Figure~(5.a) shows the convergence factor $\xi^{(k)}$ of the solution error
as a function of time.
The solution error obtained at the end of the runs
after the pulse has traveled 3 width-lengths ($t \approx 60$)
is plotted in Fig.~(5.b).
The solution errors suggest a convergence rate $O(h^{1.39})$.
This convergence rate implies that $\xi^{(k)} \rightarrow 2.62$
as $h \rightarrow 0$, which is consistent with the behavior of
$\xi^{(k)}$ in Fig.~(5.a).

Finally, we considered again the case in which the speed of sound
has a jump given by (\ref{sound_jump}); the speed of the particles was
set to $v=1$, and the source term $-\beta\,v^2\,\dt\,\nabla^2\widetilde\phi$
of the previous test was not included.
Figure~(6) shows snapshots at the same time intervals as in Fig.~(4);
for comparison, in last plot of Fig.~(6), solution differences between
the moving and fixed particle cases are presented.

\vspace{24pt}
\thesection{\centerline{\bf 5. SPI AND BURGERS EQUATION}}
\setcounter{section}{5}
\setcounter{equation}{0}
\vspace{12pt}

It is important to study the stability behavior of
SPI discretization for nonlinear equations. For this purpose,
we consider now Burgers equation, which reads
\begin{equation}
\label{burgers}
\partial_t\phi+\phi\,\partial_x\phi=\alpha\,\partial^2_x\phi.
\end{equation}

Due to the parabolic nature of Burgers equation,
the temporal integration used in previous sections
to handle the wave equation becomes unstable when
applied to this case.
For the sake of simplicity, we will consider first the
advection-diffusion equation which is obtained from
Burgers equation after replacing $\phi\,\partial_x\phi$ by $v\,\partial_x\phi$.
Here again one needs to be careful with the correct centering of each of
the terms in the equation. However, correct centering does not guarantee
stability; it is well known that a
centered-space and centered-time discretization of the
advection-diffusion equation, although second-order accurate,
is unstable. A simple cure, at the expense of sacrificing second-order
accuracy, is obtained by using forward-time, centered-space scheme.
We are interested, nonetheless, in obtaining a scheme as close as possible
to second-order accuracy and achieving, at the same time, stability.

Our approach to solve the advective-diffusion equation is analogous
to the FD scheme that combines leapfrog advection differences
($O(\dt^2,\dx^2)$)
with forward-time, centered-space differencing of diffusion terms
($O(\dt,\dx^2)$) (Roache 1985). For $\alpha \propto O(\dt)$, this method is
($O(\dt^2,\dx^2)$).
Under this method, the advection diffusion equation has a SPI
approximation
\begin{equation}
\label{adv-dif-spi}
\frac{\phi_a^{n+1} - \phi_a^{n-1}}{\dt}=
-v_a\,\sum_{b=1}^N \frac{\phi_b^n}{n_b}\partial_{x_a} W_{ab}
+\alpha\sum_{b=1}^N \frac{\phi_b^{n-1}}{n_b}\partial^2_{x_a} W_{ab}.
\end{equation}

As with the previous section, to study the convergence rate,
we have added a source term to (\ref{adv-dif-spi}) of the form
$-\alpha\,\partial^2_x\widetilde\phi$, so an analytic solution to this modified
advection-diffusion equation is given by the gaussian pulse
(\ref{pulse_form1}).
Figure~(7.a) shows the convergence factor $\xi^{(k)}$ of the solution error.
In addition, the solution errors after the pulse has
traversed three times its original thickness are
plotted as a function of the particle
separation in Fig.~(7.b) for $v=1$ and $\alpha = 1$.
These solution errors imply a $O(h^{1.95})$ scheme
and $\xi^{(k)} \rightarrow 4$ as $h \rightarrow 0$.

A similar procedure to that used for the advection-diffusion equation
was applied to Burgers equation.  The stable
SPI representation of Burgers equation is given by
\begin{equation}
\label{burgers_spi}
\frac{\phi_a^{n+1} - \phi_a^{n-1}}{\dt}=
-\phi_a^n\,\sum_{b=1}^N \frac{\phi_b^n}{n_b}\partial_{x_a} W_{ab}
+\alpha\sum_{b=1}^N \frac{\phi_b^{n-1}}{n_b}\partial^2_{x_a} W_{ab}.
\end{equation}
Here again, to test the order of the method, we added a term
$\widetilde\phi\,\partial_x\widetilde\phi-\alpha\,\partial^2_x\widetilde\phi$,
so the equation has an analytic solution $\widetilde\phi$
given by the pulse (\ref{pulse_form1}).
The convergence factor $\xi^{(k)}$ of the solution error
is plotted in Fig~(8.a) and the final solution error in Fig.~(8.b)
for $\alpha = 0.2$.
We find that the SPI discretization (\ref{burgers_spi}) to Burgers
equation gives solution error of $O(h^{1.98})$.
Finally, in Figure~(9) we show the solution to Burgers equation,
without the source terms
$\widetilde\phi\,\partial_x\widetilde\phi-\alpha\,\partial^2_x\widetilde\phi$,
for an initially gaussian pulse.
As expected, the pulse develops shock-like features due to the non-linear
term and also diffuses because of the dissipation term.

\vspace{24pt}
\thesection{\centerline{\bf 6. CONCLUSIONS}}
\setcounter{section}{6}
\setcounter{equation}{0}
\vspace{12pt}

We have presented a method, smoothed particle interpolations (SPI),
based on SPH techniques to solve
hyperbolic and parabolic equations of non-hydrodynamic nature.
The method was tested solving the one-dimensional
wave equation with inhomogeneous sound speed and Burgers equation.
We showed that the intrinsic adaptive nature of SPH is directly
carried over. We also provided a prescription for handling advection-type
terms that arise when particles are free to move or when dealing with
Burgers equation. Furthermore, we showed the feasibility of obtaining
$O(h^2)$ schemes under SPI discretizations.

A possible drawback of this approach is its particle requirements
compared with the mesh-point cost in finite difference methods.
We have shown and tested that, for uniformly spaced particles,
SPI required a minimum of three smoothing
particles for each grid point in FD in order to achieve equivalent
absolute solution errors.

In principle, generalization of this work to two or
three dimensional systems
should be straightforward (Laguna 1994).
After all, one of the reasons for SPH's popularity has been its
simplicity for implementation in higher dimensions.
Moreover, all the technology developed in N-body methods
and standard SPH for finding neighboring particles
can be directly utilized.

\vspace{24pt}
\centerline{\bf ACKNOWLEDGMENTS}
\vspace{12pt}
I thank Richard Matzner and Matt Choptuik for numerous discussions
and helpful suggestions.
Work supported in part by the NASA
(at Los Alamos National Laboratory),
NSF Young Investigator award PHY-9357219,
and NSF grant PHY-9309834.

\vfill\eject
\def\prl#1#2#3{ #1, {\it Phys. Rev. Lett.}, {\bf #2}, #3}
\def\prd#1#2#3{ #1, {\it Phys. Rev. D}, {\bf #2}, #3}
\def\jcp#1#2#3{ #1, {\it J. Comput. Phys.}, {\bf #2}, #3}
\def\apj#1#2#3{ #1, {\it Astrophys. J.}, {\bf #2}, #3}
\def\sia#1#2#3{ #1, {\it SIAM J. Sci. Statist. Comput.}, {\bf #2}, #3}
\def\aa#1#2#3{ #1, {\it Astron. Astrophys.}, {\bf #2}, #3}
\def\hangpara{\par\hangindent 25 pt\noindent}

\vspace{24pt}
\centerline{\bf REFERENCES}
\vspace{12pt}

\hangpara
Benz, W. 1990, in {\it Numerical Modeling of Stellar
Pulsation: Problems \& Prospects}, ed. J.R. Buchler,
(Dordrecht: Kluwer Academic), p269

\hangpara
Berger, M.J., \& Oliger, J. \jcp{1984}{53}{484}

\hangpara
Choptuik, M. (1986) Ph.D. thesis, University of British Columbia

\hangpara
Choptuik, M. \prd{1991}{44}{3124}

\hangpara
Choptuik, M. \prl{1993}{70}{9}

\hangpara
Davies, M.B., Benz, W. \& Hills, J. \apj{1991}{381}{449}

\hangpara
Davis, S., \& Flaherty, J. \sia{1982}{3}{6},

\hangpara
Evans, C. R. \& Kochanek, C. S. \apj{1989}{346}{L13}

\hangpara
Hawley, J. F., Smarr, L. L., \& Wilson, J. R. \apj{1984}{277}{296}

\hangpara
Hernquist, L., \& Katz, N. \apj{1989}{70}{419}

\hangpara
Laguna, P., Miller, W.A., \& Zurek, W. H. \apj{1993}{404}{678}

\hangpara
Laguna, P., Miller, W.A., Zurek, W.H., \& Davies, M.B.
\apj{1993}{410}{L83}

\hangpara
Laguna, P. 1993, in preparation

\hangpara
Mann, P.J. \jcp{1993}{107}{188}.

\hangpara
Monaghan, J. J. \jcp{1989}{82}{1}

\hangpara
Monaghan, J. J., \& Lattanzio, J. C. \aa{1985}{149}{135}

\hangpara
O'Rourke, P.J., \& Amsden, A.A. 1986, Los Alamos National Laboratory
report LA-10849-MS

\hangpara
Roache, P.J. 1985 {\it Computational Fluid Dynamics},
(Albuquerque: Hermosa Publishers)

\hangpara
Stellingwerf, R.F., \& Peterkin, R.E. 1989,
{\it Mission Research Corporation} report MRC/ABQ-R-1248

\eject

\vspace{24pt}
\centerline{\bf FIGURE CAPTIONS}
\vspace{12pt}

\begin{description}
\item[Fig.1] Gaussian kernel with compact support radius
             $u_o = 5$. The compact support radius is defined
             as $u_o = N_{1/2}\dx/h$ where $h$ is the smoothing
             length, $\dx$ the particle interseparation and
             $N_{1/2} = 10$ the number of neighbors of a particle
             at each of its {\it sides}. In the figure, the
             particle positions are denoted by open circles.

\item[Fig.2] Smoothed values errors in the representation
             of $f(x) = \sin(x)$ and its
             smoothed first and second derivatives for
             $\dx^{(k)} = 2^{-k} \dx_o, k$=1(circle),2(cross),
             3(triangle),4(square)
             and $h^{(l)} = 2^{-l} h_o,\, l = 1,2,3,4.$

\item[Fig.3] Results from a gaussian pulse propagating
             in a medium with uniform sound speed $c=1$.
             The particles were kept fixed throughout the evolution.
             (a) shows solution errors as a function of the
             particle interseparation.
             (b) shows the differences between
             SPI and FD solution errors.
             The errors were obtained after a pulse displacement of
             four initial gaussian width-lengths.

\item[Fig.4] Propagation of a gaussian pulse through a medium
             in which the sound speed has a jump
             at the origin given by Eq.~(\ref{sound_jump}).
             The particles were kept fixed throughout the evolution.
             Bottom plot shows the difference between the SPI
             and FD solution in last snapshot.

\item[Fig.5] Results from propagating a gaussian pulse
             in a medium with uniform sound speed $c~=~1$.
             The particles moved with speed $v=1$.
             (a) shows the convergence factor $\xi^{(k)}$ of the solution
error.
             (b) shows the solution errors as a function of the
             particle interseparation at a pulse displacement of
             four pulse width-lengths.

\item[Fig.6] Same as in Fig.~(3) but for particles moving with
             speed $v=1$. The plot at the bottom in this case shows
             the differences of the solutions between the fixed and moving
             particle cases at the end of the run.

\item[Fig.7] (a) Convergence factor $\xi^{(k)}$ of the solution error
             for the advective-diffusion equation.
             A source term of the form
             $-\alpha\,\partial^2_x\widetilde\phi$ was added to the equation
             in order to have a gaussian pulse as analytic solution.
             (b) Solution errors as a function of the
             particle interseparation.

\item[Fig.8] Same as in Fig.~(6) but for Burgers equation with a source
             term $\widetilde\phi\,\partial_x\widetilde\phi
             -\alpha\,\partial^2_x\widetilde\phi$.

\item[Fig.9] Solution to Burgers equation of an initial gaussian pulse
             with $\alpha = 0.2$.

\end{description}

\end{document}